\documentstyle[preprint,aps,floats,psfig]{revtex}
\newcommand {\dgbb} {\Delta\Gamma_{\gamma bb}^{\rm NP}}
\newcommand {\dzbb} {\Delta\Gamma_{Zbb}^{\rm NP}}
\newcommand {\dgbbqsq} {\Delta\Gamma_{\gamma bb}^{\rm NP}(q^2)}
\newcommand {\dzbbqsq} {\Delta\Gamma_{Zbb}^{\rm NP}(q^2)}

\newcommand {\barssq}	{\overline{s}^2}

\newcommand {\bargzsq}	{\overline{g}_Z^2}
\newcommand {\bargwsq}	{\overline{g}_W^2} 

\newcommand {\baralphaqsq}	{\overline{\alpha}(q^2)}

\newcommand {\barssqqsq}	{\overline{s}^2(q^2)}

\newcommand {\bargzsqqsq}	{\overline{g}_Z^2(q^2)}
\newcommand {\bargwsqqsq}	{\overline{g}_W^2(q^2)}

\newcommand {\qsq} {q^2}

\newcommand {\barf}      {\overline{f}}

\newcommand {\bardeltab} {\overline{\delta}_b}

\newcommand {\hate}	{\hat{e}}
\newcommand {\hatg}	{\hat{g}}
\newcommand {\hats}	{\hat{s}}
\newcommand {\hatc}	{\hat{c}}
\newcommand {\hatgz}	{\hat{g}_Z}
 
\newcommand {\hatalpha}	{\hat{\alpha}} 
\newcommand {\hatesq}	{\hat{e}^2}
\newcommand {\hatgsq}	{\hat{g}^2}
\newcommand {\hatssq}	{\hat{s}^2}
\newcommand {\hatcsq}	{\hat{c}^2}
\newcommand {\hatgzsq}	{\hat{g}_Z^2}

\newcommand {\mwsq}  {m_W^2}
\newcommand {\mzsq}  {m_Z^2}
\newcommand {\Lamsq} {\Lambda^2}

\newcommand {\fwww}    {f_{WWW}}
\newcommand {\fw}      {f_{W}}
\newcommand {\fb}      {f_{B}}

\newcommand {\owww}    {{\cal O}_{WWW}}
\newcommand {\ow}      {{\cal O}_{W}}
\newcommand {\ob}      {{\cal O}_{B}}

\newcommand {\oefour}  {${\cal O}(E^4)\,$}
\newcommand {\oesix}   {${\cal O}(E^6)\,$}
\newcommand {\oeeight} {${\cal O}(E^8)\,$}

\newcommand {\call}  {{\cal L}}

\newcommand {\kapgam}  {\kappa_\gamma}
\newcommand {\kapz}    {\kappa_Z}
\newcommand {\lamgam}  {\lambda_\gamma}
\newcommand {\lamz}    {\lambda_Z}
\newcommand {\gonegam} {g_1^\gamma}
\newcommand {\gonez}   {g_1^Z}

\newcommand {\dkapgam}  {\Delta\kappa_\gamma}
\newcommand {\dkapz}    {\Delta\kappa_Z}

\newcommand {\dgonegam} {\Delta g_1^\gamma}
\newcommand {\dgonez}   {\Delta g_1^Z}

\newcommand {\via} {{\em via}}
\newcommand {\etc} {{\em etc.}}
\newcommand {\ie} {{\em i.e.}}

\draft

\preprint{\vbox{\baselineskip14pt
\hbox{\bf KEK-TH-531}
\hbox{\bf KEK Preprint 97-107}
\hbox{\bf hep-ph/9708218}
\hbox{July 1997}}}
\title{Contributions of LEP1.5, LEP2 and linear-collider data to indirect 
constraints on non-Abelian gauge-boson couplings}
\author{R.~Szalapski}
\address{Theory Group, KEK, Tsukuba, Ibaraki 305, Japan}

\begin{document}
\maketitle 

\begin{abstract}
It is possible to place direct constraints on $WW\gamma$ and $WWZ$
couplings by studying their tree-level contributions to the process 
$e^+e^-\rightarrow W^+W^-$.\/  However, these couplings also contribute at the 
loop level to $e^+e^-\rightarrow f\barf$ processes where $f$ is any 
Standard-Model fermion.   In this paper the available LEP1.5 and LEP2 data,
the anticipated LEP2 data and possible linear collider data for these 
latter processes is combined with low-energy and $Z$-pole data to place indirect 
constraints on nonstandard $WW\gamma$ and $WWZ$ couplings.  The direct and 
indirect constraints are then compared.
An effective 
Lagrangian is used to describe the new physics.  In order that the implications 
of this analysis are as broad as possible, both the light-Higgs scenario, 
described by an effective Lagrangian with a linear realization of the 
symmetry-breaking sector, and the strongly interacting scenario, described by 
the electroweak chiral Lagrangian, are considered.  
\end{abstract}

\newpage 

\section{Introduction and parameterizations}
Non-Abelian gauge-boson couplings are an essential and fascinating aspect 
of the Standard Model (SM).   With the aim of verifying the SM or detecting 
new physics it is important to measure such couplings.  The basic strategy 
is to introduce and then measure the most general couplings allowed under a 
particular set of physical assumptions.  If, for example, one requires the new 
physics to be invariant under U(1)$_{\rm em}$, then the most general $WW\gamma$ 
and $WWZ$ couplings allowed are parameterized by the effective 
Lagrangian\cite{hpzh87}
\begin{eqnarray}
\nonumber
\lefteqn{\call_{WWV} =}\\&&  - i g_{WWV} \Bigg\{ 
g_1^V \Big( W^+_{\mu\nu} W^{- \, \mu} V^{\nu} 
  - W^+_{\mu} V_{\nu} W^{- \, \mu\nu} \Big) 
+ \kappa_V W_\mu^+ W_\nu^- V^{\mu\nu}
+ \frac{\lambda_V}{\mwsq} W^+_{\mu\nu} W^{- \, \nu\rho} V_\rho^{\; \mu}
 \Bigg\}
\;,\label{lagr-phenom}
\end{eqnarray}
where $V=Z,\gamma$, the overall coupling constants are $g_{WW\gamma} = \hate$ 
and $g_{WWZ} = \hatgz\hatcsq$.  \footnote{The `hatted' couplings are the 
$\overline{\rm MS}$ couplings which satisfy the tree-level relations 
$\hate = \hatg\hats = \hatgz\hats\hatc$ and $\hatesq = 4\pi\hatalpha$;  $\hatg$ 
is the SU(2) coupling, $\hats$ and $\hatc$ are the sine and cosine of the weak 
mixing angle, the strength of the photon coupling is given by $\hate$, and 
$\hatg^\prime = \hatg\hats/\hatc$ is the U(1) coupling.} Here the field-strength 
tensors include only the Abelian parts, \ie\/
$W^{\mu\nu} = \partial^\mu W^\nu - \partial^\nu W^\mu$
and $V^{\mu\nu} = \partial^\mu V^\nu - \partial^\nu V^\mu$.  
Eqn.~(\ref{lagr-phenom}) has been truncated to include only terms which 
separately conserve charge conjugation (C) and parity (P).  While other 
operators exist, they shall be irrelevant in the ensuing discussion.  Notice 
that $\kapgam$, $\kapz$, $\gonegam$ and $\gonez$ are couplings associated with 
energy-dimension-four (\oefour) operators while $\lamgam$ and $\lamz$
coincide with energy-dimension-six (\oesix) operators.  The effect of 
including additional operators (with higher energy dimension) is equivalent to 
a running of the couplings, \ie\/ $\kapgam = \kapgam(\qsq)$, $\kapz = 
\kapz(\qsq)$, \etc \/\footnote{In the SM, at the tree level, $\kapgam = \kapz = 
\gonegam = \gonez = 1$ and $\lamgam = \lamz = 0$.  At higher orders 
$\gonegam(\qsq = 0) = 1$ due to a U(1)$_{\rm em}$ Ward identity.}

The next logical step is to impose the full symmetries of the SM; considering 
only electroweak interactions this implies imposing an SU(2)$\times$U(1) 
symmetry spontaneously broken to U(1)$_{\rm em}$.  In order to proceed it is 
necessary to choose the method of spontaneous symmetry breaking (SSB); is 
symmetry breaking linearly realized? \ie\/ 
Is there a physical Higgs scalar? Or is the nonlinear realization 
appropriate?  First discussing the linear realization, one may extend the 
SM Lagrangian\cite{bw86andbs83andllr86} according to
\begin{equation}
\label{leff-lr}
\call_{\rm eff}^{\rm linear} = \call_{\rm SM} 
       + \frac{\fw}{\Lamsq} {\ow} 
       + \frac{\fb}{\Lamsq} {\ob} 
       + \frac{\fwww}{\Lamsq} {\owww} 
       + \cdots \;.
\end{equation}
Here $\call_{\rm SM}$ is the usual SM Lagrangian including an SU(2) doublet 
Higgs field, $\Phi$.   The operators ${\cal O}_i$ are \oesix operators; each is
accompanied by a dimensionless coupling $f_i$ and suppressed by a factor 
$\Lambda^2$, where $\Lambda$ is the scale associated with the new physics.
Additional operators exist which are either stringently constrained by the 
current data or are irrelevant to $WW\gamma$ and $WWZ$ couplings.\cite{hisz}  
As the measurements of $WW\gamma$ and $WWZ$ couplings improve it will be 
necessary to include additional operators, but currently these three are 
sufficient\cite{hhis96}.  There are no \oesix operators in the light-Higgs 
scenario which conserve CP without separately conserving C and P.  For explicit 
expressions of the operators and further details see 
Refs.~\cite{hisz,hhis96,hms95}.

From Eqn.~(\ref{leff-lr}) it follows that\cite{hisz}:
\begin{mathletters}
\label{hisz-relations}
\begin{eqnarray}
\label{g1gam-hisz}
\gonegam(\qsq) & = & 1 \;,
\\
\label{g1z-hisz}
\gonez(\qsq) & = & 1 + \frac{1}{2}\frac{\mzsq}{\Lamsq}\fw \;, 
\\
\label{kappagamma-hisz}
\kapgam(\qsq) & = & 1 + \frac{1}{2}\frac{\mwsq}{\Lamsq}\Big(\fw + \fb\Big) \;, 
\\
\label{kappaz-hisz}
\kapz(\qsq) & = & 1 + \frac{1}{2}\frac{\mzsq}{\Lamsq}
  \Big(\hatcsq\fw - \hatssq\fb\Big)\;, 
\\
\label{lambda-hisz}
\lamgam(\qsq) = \lamz(\qsq) & = & \frac{3}{2}\hatgsq\frac{\mwsq}{\Lamsq}\fwww \;.
\end{eqnarray}
\end{mathletters}
Of six couplings in Eqn.~(\ref{lagr-phenom}), only three are independent.
In particular, from the set $\{\kapgam,\kapz,\gonez\}$, only two are independent.
These relationships are broken by the inclusion of \oeeight 
operators\cite{hisz}.

In the nonlinear realization, employing the notation of 
Ref.~\cite{lon80andlon81andaw93}, the relevant Lagrangian becomes 
\begin{equation}
\label{leff-nlr}
\call_{\rm eff}^{\rm nlr} = \call_{\rm SM}^{\rm nlr} 
	+ \call_{2} 
	+ \call_{3} 
	+ \call_{9} 
	+ \cdots \;,
\end{equation}
where the superscript `nlr' denotes `nonlinear realization'.  Again only 
those operators which are not stringently constrained but are relevant to 
$WW\gamma$ and $WWZ$ couplings are included.  The 
first term is the SM Lagrangian without a physical Higgs boson; it is hence 
nonrenormalizable.  The terms $\call_{2}$, $\call_{3}$ and $\call_{9}$
introduce the dimensionless parameters $\alpha_2$, $\alpha_3$ and $\alpha_9$
respectively, each multiplying an \oefour operator; for further details see
Refs.~\cite{hhis96,lon80andlon81andaw93}.  All three separately conserve
C and P, and only $\call_{9}$ breaks the custodial SU(2)$_{\rm C}$ symmetry.
Note that there is one CP-conserving operator, $\call_{11}$, that conserves 
neither C nor P.  This additional operator contributes to a P-violating $WWZ$ 
coupling, but it is not easily incorporated in the current analysis; it has been 
discussed elsewhere\cite{dv94}.

From Eqn.~(\ref{leff-nlr}) one may obtain a set of relationships similar to 
those of Eqn.~(\ref{hisz-relations}).  In particular\cite{hhis96,dv94,fer93},
\begin{mathletters}
\label{nl-hisz-relations}
\begin{eqnarray}
\label{nl-g1z-hisz}
\gonez(\qsq) & = & 1 + \hatgzsq\alpha_3 \;,
\\
\label{nl-kappagamma-hisz}
\kapgam(\qsq) & = & 1 + \hatgsq \Big( \alpha_2 + \alpha_3 + \alpha_9 \Big)\;, 
\\
\label{nl-kappaz-hisz}
\kapz(\qsq) & = & 1 + \hatgzsq\Big( 
-\hatssq\alpha_2  + \hatcsq\alpha_3 + \hatcsq\alpha_9 \Big)  \;, \\
\label{nl-lambda-hisz}
\lamgam(\qsq) & = & \lamz(\qsq)   \approx  0 \;.
\end{eqnarray}
\end{mathletters}
The couplings $\lamgam$ and $\lamz$ are equal only in the sense that that 
they are both zero, and at higher orders it is expected that these two parameters
will be nonzero and unequal.  Notice that, in the set $\{\kapgam,\kapz,\gonez\}$,
all three are independent.  If the new physics is SU(2)$_{\rm C}$
invariant, the contribution of $\alpha_9$ may be dropped.  Then the 
relations among $\{\kapgam,\kapz,\gonez\}$ are the same as those of 
Eqns.~(\ref{g1z-hisz})-(\ref{kappaz-hisz}).  This is clarified by making the 
identifications $\hatgsq \alpha_2 = \frac{1}{2}\frac{\mwsq}{\Lamsq}\fb$ 
and $\hatgsq \alpha_3 = \frac{1}{2}\frac{\mwsq}{\Lamsq}\fw$.
 

\section{Contributions to observables}

To place constraints upon $\fw$, $\fb$ and $\fwww$, it is necessary to 
calculate the contributions of the operators $\ow$, $\ob$ and $\owww$
to amplitudes with four external light fermions.  To constrain $\alpha_2$, 
$\alpha_3$ and $\alpha_9$ it is necessary to repeat the procedure for 
$\call_2$, $\call_3$ and $\call_9$.  It is particularly convenient to 
organise the overall calculation according to Ref.~\cite{hhkm94}.  The 
propagator corrections and the pinch-portions of the vertex corrections are 
absorbed into the charge form factors $\baralphaqsq$, $\bargzsqqsq$, $\barssqqsq$
and $\bargwsqqsq$.  (The pinch 
technique\cite{kl89,cp89andpap90,ds92anddks93,pinch} renders the 
propagator and vertex corrections separately gauge invariant.)
With the exception of any non-SM vertex corrections which 
must be added explicitly, the SM vertex and box corrections are employed.  The 
two-point-functions in the linear realization were calculated in 
Ref.~\cite{hisz}; for the nonlinear realization, see 
Refs.~\cite{hisz,dv95,ads97}.  Expressions for the `barred' charges appear 
in Ref.~\cite{ads97}; being rather lengthy, they are not reproduced here.

\begin{figure}[tb]
\begin{center}
\leavevmode\psfig{file=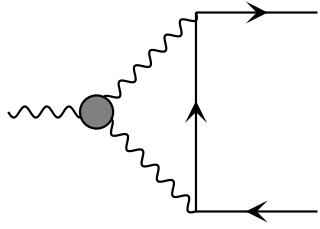,angle=0,height=3cm,silent=0}
\end{center}
\caption{Non-SM corrections to the $Zbb$ and $\gamma bb$ vertex. The new physics 
enters through the `blob' vertex.}
\label{fig-vff} 
\end{figure}

Non-SM contributions to the $Zbb$ vertex were presented in Ref.~\cite{ads97}.
In the current context it is necessary to consider non-SM contributions to 
the $\gamma bb$ vertex as well.  See Fig.~\ref{fig-vff}.
In order to avoid a 
conflict with existing notation found in the references, but at the same time 
wishing to avoid a lengthy discussion of vertex corrections, two new parameters, 
$\dgbbqsq$ and $\dzbbqsq$ are introduced.  (The superscript `NP' indicates new 
physics.)  They are defined according to how they modify the SM Feynman rules:
\begin{mathletters}
\label{delta-vbb}
\begin{eqnarray}
\label{delta-gbb}
& -i \hate\gamma^\mu\bigg\{ Q_b P_+ + \Big[ \dgbbqsq I_3 + Q_b\Big]P_-\bigg\}
&\;,\\
\label{delta-zbb}
& -i \hatgz\gamma^\mu\bigg\{ -\hatssq Q_b P_+ + 
\Big[ \big(1 + \dzbbqsq\big) I_3 - \hatssq Q_b\Big]P_
-\bigg\}& \;,
\end{eqnarray}
\end{mathletters}
for the $\gamma bb$ and $Zbb$ vertices respectively.  The one-loop contributions
of the SM are not shown.  The projection operators are defined by 
$P_\pm = (1\pm\gamma_5)/2$, and $Q_b = -1/3$ and $I_3 = -1/2$ are the charge and 
weak-isospin quantum numbers of the b quark.  Also, $\dgbb(0) = 0$ by Ward 
identities\cite{hhkm94} while $\dzbbqsq$ may contain a constant term.  Both of
these form factors receive contributions from $\ow$ and $\ob$ through a 
new-physics contribution to the $WW\gamma$ and $WWZ$ couplings as depicted 
in Fig.~\ref{fig-vff}.  (They receive corrections from $\call_2$, $\call_3$
and $\call_9$ in the nonlinear realization of SSB.) Recall that the pinch term 
is removed from the vertex correction form factor but included in the barred 
charges.  An explicit calculation yields
\begin{mathletters}
\label{delta-vbb-gen}
\begin{eqnarray}
\label{delta-gbb-gen}
\dgbbqsq & = & 
-\frac{\hatgsq}{64\pi^2}\frac{m_t^2}{\mwsq}\bigg\{\frac{\qsq}{\mwsq}\dkapgam
(\qsq) + 6 \dgonegam(\qsq) \bigg\}\log\bigg( \frac{\Lambda^2}{\mzsq} \bigg)
\;,\\
\label{delta-zbb-gen}
\dzbbqsq & = & 
-\frac{\hatgsq}{64\pi^2}\frac{m_t^2}{\mzsq}\bigg\{\frac{\qsq}{\mwsq}\dkapz
(\qsq) + 6 \dgonez(\qsq) \bigg\}\log\bigg( \frac{\Lambda^2}{\mzsq} \bigg)
\;,
\end{eqnarray}
\end{mathletters}
where $\dkapgam = \kapgam-1$, $\dkapz = \kapz-1$, 
$\dgonegam = \gonegam-1$ and $\dgonez = \gonez-1$.  Recall that 
$\gonegam(0) = 1$.  An expression equivalent to Eqn.~(\ref{delta-zbb-gen}) was 
presented in Ref.~\cite{ads97}.  Replacing a $WWZ$ vertex in Fig.~\ref{fig-vff}
with a $WW\gamma$ vertex introduces a factor of $\hats/\hatc$ and requires the 
replacements $\dkapz \rightarrow \dkapgam$ and 
$\dgonez \rightarrow \dgonegam$.  Taking into the account different 
overall factors in the definitions of Eqn.~(\ref{delta-vbb}), the expression 
for $\dgbbqsq$ is obtained from the expression 
for $\dzbbqsq$.  Explicit expressions for $\kapgam$, 
$\kapz$, $\gonegam$ and $\gonez$ from Eqns.~(\ref{hisz-relations}) and 
Eqns.~(\ref{nl-hisz-relations}) may be used to obtain the appropriate 
form factors for the linear and nonlinear realizations of SSB respectively.


\section{The data}

From the recent analysis of Ref.~\cite{hhm97}, the data for low-energy and 
$Z$-pole measurements is nicely summarized as measurements of the various charge 
form factors.  That global analysis yielded, for measurements on the $Z$-pole,
\begin{equation}
\label{z-pole-data}
\left. \begin{array}{l}
\bargzsq(\mzsq) = 0.55557 - 0.00042
\frac{\alpha_s + 1.54 \bardeltab(\mzsq) - 0.1065}{0.0038} \pm 0.00061 
\\
\barssq(\mzsq) = 0.23065 + 0.00003
\frac{\alpha_s + 1.54 \bardeltab(\mzsq) - 0.1065}{0.0038} \pm 0.00024
\end{array}\right\}\makebox[.2cm]{}
\rho_{\rm corr} = 0.24 \;,
\end{equation}
where $\bardeltab(\mzsq) = \overline{\delta}_{b\; \rm SM}(\mzsq)
+ \dzbb(\mzsq)$; the precise definition of 
$\bardeltab(\qsq)$ is found in Ref.~\cite{hhkm94}.  Combining the 
$W$-boson mass measurement ($m_W = 80.356\pm 0.125$~GeV) with the input 
parameter $G_F$,
\begin{eqnarray}
\label{w-data}
\bargwsq(0) = 0.4237 \pm 0.0013\;.
\end{eqnarray}
And finally, from the low-energy data,
\begin{equation}
\label{le-data}
\left. \begin{array}{l}
\bargzsq(0) = 0.5441 \pm 0.0029 \\
\barssq(0) = 0.2362 \pm 0.0044
\end{array}\right\}\makebox[0.2cm]{} \rho_{\rm corr} = 0.70 \;.
\end{equation}
The implications of this data for non-Abelian gauge-boson couplings was 
considered in Ref.~\cite{ads97}.  Note that the above data is insensitive to 
$\dgbbqsq$; because $\dgbb(0) = 0$, there is a 
negligible contribution at low energies, and on the $Z$-pole the effects of 
photon exchange are negligible.  Also notice that there are no constraints on 
$\baralphaqsq$ and $\bargwsqqsq$ away from $\qsq = 0$.

Data is now available from measurements at LEP1.5 and LEP2.  This has 
implications for the indirect constraints on new physics.  In particular, there 
is now sensitivity to $\dgbbqsq$ and $\baralphaqsq$.  Additionally some of the 
contributions to the charge form factors run with $\qsq$, hence measurements at 
different center of mass (CM) energies constrain different linear combinations 
of the parameters.  Some of the contributions are enhanced as the CM energy 
increases, but unfortunately there is also a loss in statistics.

The three LEP detector collaborations have published measurements for 
$e^+e^- \rightarrow f\barf$ where $f$ is a typical fermion for
LEP1.5 energies of 130-140~GeV\cite{aleph-plb378,l3-plb370,opal-plb376}
and LEP2 energies of 161-172~GeV\cite{opal-plb391,l3-ppe9752}.  The 
relevant measurements are summarised in Table~\ref{table-lep2}.
\begin{table}[tbh]
\begin{tabular}{|l||c|c|c|c|c|c|c|c|}
Detector & $\sqrt{s}$(GeV) & $\cal L$(pb$^{-1}$) & $\sigma_{\rm had}$(pb) &
$\sigma_{\mu^+\mu^-}$(pb) & $\sigma_{\tau^+\tau^-}$(pb) & 
$A_{\rm fb}^{\mu^+\mu^-}$ & $A_{\rm fb}^{\tau^+\tau^-}$ & $R_b$
\\ \hline\hline
ALEPH & 130 & 2.9 & 74.2$\pm$6.2 & 9.6$\pm$1.9 & 11.8$\pm$2.3 &
0.65$^{+0.16}_{-0.23}$ & 0.91$^{+0.08}_{-0.20}$ & 
\\ \hline
ALEPH & 136 & 2.9 & 57.4$\pm$4.8 & 7.1$\pm$1.7 & 5.8$\pm$1.7 & 
0.53$^{+0.22}_{-0.29}$ & 0.70$^{+0.18}_{-0.42}$ &
\\ \hline\hline
L3 & 130.3 & 2.7 & 81.8$\pm$6.4 & 7.7$\pm$1.8 & 10.4$\pm$2.8 & 
0.83$^{+0.16}_{-0.22}$ & 0.65$^{+0.15}_{-0.25}$ &
\\ \hline
L3 & 136.3 & 2.3 & 70.5$\pm$6.2 & 6.1$\pm$1.7 & 9.4$\pm$2.8 & 
0.92$^{+0.08}_{-0.27}$ & 0.98$^{+0.02}_{-0.23}$ &
\\ \hline
L3 & 140.2 & 0.05 & 67$\pm$47 & &&&&
\\ \hline
L3 & 161.3 & 10.9 & 37.3$\pm$2.3 & 4.59$\pm$0.86 & 4.6$\pm$1.1 & 
0.59$^{+0.14}_{-0.18}$ & 0.97$^{+0.10}_{-0.25}$ &
\\ \hline
L3 & 170.3 & 1.0 & 39.5$\pm$7.5 & &&&&
\\ \hline
L3 & 172 & 10.2 & 28.2$\pm$2.3 & 3.60$\pm$0.76 & 4.3$\pm$1.1 & 
0.31$^{+0.19}_{-0.22}$ & 0.18$^{+0.27}_{-0.29}$ &
\\ \hline\hline
OPAL & 130.26 & 2.7 & 66$\pm$6 & 9.5$\pm$1.9 & 6.0$\pm$2.0 &&&
\\ \hline
OPAL & 136.23 & 2.6 & 60$\pm$5 & 11.6$\pm$2.1 & 7.6$\pm$2.3 &&&
\\ \hline
OPAL & 133 &&&&& 0.65$\pm$0.12 & 0.31$\pm$0.16 & 0.216$\pm$0.036 
\\ \hline
OPAL & 140 & 0.04 & 50$\pm$36 &&&&&
\\ \hline
OPAL & 161.3 & 10.0 & 35.3$\pm$2.1 & 4.6$\pm$0.7 & 6.7$\pm$1.0 & 0.49$\pm$0.14 &
0.52$\pm$0.14 & 0.141$\pm$0.031
\end{tabular}
\caption{LEP1.5 and LEP2 data from the three LEP detector collaborations.  
Statistical and systematic errors, where reported separately, have been combined 
in quadrature.  OPAL entries at 133~GeV are actually the average of measurements 
at 130~GeV and 136~GeV.}
\label{table-lep2}
\end{table}
Notice that the table is somewhat incomplete.  ALEPH measurements at LEP2 
energies are not yet available, and only L3 has reported measurements above 
170~GeV.   Only OPAL reports on measurements specific to b quarks.  The results 
for OPAL at 133~GeV were obtained by combining data at 130~GeV and 
136~GeV; no data was actually taken at 133~GeV.  Whenever statistical and 
systematic errors were reported separately, they have been combined in quadrature
before being entered into the table.  The various table entries are treated as 
separate and independent measurements; no attempt is made to directly combine 
the results of the different collaborations.  

A significant portion of the events above the $Z$ peak are ``radiative return''
events.  That is, a photon is radiated which reduces the effective CM energy to 
$\sqrt{\hat{s}}\approx m_Z$.  Except for the appearance of the extra radiation, 
such events are characteristically the same as LEP1 events on $Z$-pole.  In 
consideration of the enormous number of $Z$ decays accumulated at LEP1, it is 
very reasonable to neglect $e^+e^- \rightarrow f\barf\gamma$ events at 
LEP1.5 and LEP2 when the photon carries away a significant portion of the 
energy.  For this reason only the exclusive modes are reported in the table.

Additionally it is interesting to anticipate what data might 
be available in the future.  Four different data sets will be considered:
\begin{enumerate}
\item{Fit~\ref{fit-1}: All low-energy and $Z$-pole data as summarised by 
	Eqns.~(\ref{z-pole-data})-(\ref{le-data}).\label{fit-1}}
\item{Fit~\ref{fit-2}: The data from Fit~\ref{fit-1} plus the LEP1.5 and early 
      LEP2 data summarised in Table~\ref{table-lep2}.\label{fit-2}}
\item{Fit~\ref{fit-3}: The data from Fit~\ref{fit-2} plus an estimate of future 
      LEP2 data.  A luminosity of 125~pb$^{-1}$/year/detector is assumed for 2 
      detectors 
      and 3 years.  The luminosity is evenly divided between 188~GeV and 
      192~GeV.  It is also assumed that LEP2 will obtain a measurement of 
      $\Delta m_W = 50$~MeV.\label{fit-3}}
\item{Fit~\ref{fit-4}: The data from Fit~\ref{fit-3} plus 50~fb$^{-1}$ of data 
      collected at a $\sqrt{s} = 500$~GeV linear $e^+e^-$ and an assumed 
      measurement of $\Delta m_W = 20$~MeV at the TeV33.\label{fit-4}}
\end{enumerate}
For the future LEP2 data it the following observables were used: 
$\sigma (e^+e^- \rightarrow \mu^+\mu^- )$,
$\sigma (e^+e^- \rightarrow {\rm hadrons})$ and the forward-backward 
asymmetries $A^\mu_{FB}$, $A^b_{FB}$ and $A^c_{FB}$.  In every case it 
was assumed that statistical errors dominate over systematic errors.
For the linear collider the uncertainty in the luminosity measurement may 
contribute to an uncertainty in absolute rates.  Therefore, the $R_h$ is 
used in place of the hadronic cross section, and 
$\sigma (e^+e^- \rightarrow \mu^+\mu^- )$ is assigned an error of 3\%.


\section{Numerical studies and discussion}

In all numerical studies the scale of new physics is taken to be 
$\Lambda = 1$~TeV, and the renormalization scale is chosen to be 
$\mu = m_Z$. The one-sigma bounds on  $\fwww$, $\fw$ and $\fb$ are 
presented in Table~\ref{table-linear-indirect}.
\begin{table}[tbh]
\begin{tabular}{|l|c|c|c|c|c|}
  && $m_H=75$~GeV & $m_H=200$~GeV & $m_H=400$~GeV & $m_H=800$~GeV \\
\hline\hline 
$\fwww$ & Fit 1 & -19$\pm$10 & 5$\pm$10 & 25$\pm$10 & 45$\pm$10 \\
        & Fit 2 & -20$\pm$10 & 5$\pm$10 & 24$\pm$10 & 44$\pm$10 \\
        & Fit 3 & -22$\pm$10 & 4$\pm$10 & 24$\pm$10 & 45$\pm$10 \\
        & Fit 4 & -27$\pm$8  & 2$\pm$8  & 26$\pm$8  & 51$\pm$8  \\ 
\hline 
$\fw$   & Fit 1 & 1.8$\pm$3.2 & -5.0$\pm$3.7 & -7.6$\pm$4.4 & 1.9$\pm$3.8 \\
        & Fit 2 & 2.1$\pm$3.2 & -4.6$\pm$3.7 & -7.3$\pm$4.4 & 2.0$\pm$3.8 \\
        & Fit 3 & 3.0$\pm$3.0 & -4.3$\pm$3.6 & -7.4$\pm$4.4 & 3.2$\pm$3.7 \\
        & Fit 4 & 6.1$\pm$2.4 & -3.1$\pm$3.1 & -8.0$\pm$4.3 & 9.2$\pm$3.4 \\ 
\hline 
$\fb  $ & Fit 1 & -3$\pm$10   & 7.4$\pm$7.4 & -0.5$\pm$4.0 & -4.5$\pm$2.5 \\
        & Fit 2 & -2$\pm$10   & 8.2$\pm$7.4 &  0.1$\pm$4.0 & -4.1$\pm$2.5 \\
        & Fit 3 & -1$\pm$10   & 7.2$\pm$6.8 & -1.4$\pm$3.4 & -5.1$\pm$2.2 \\
        & Fit 4 & 2.8$\pm$5.7 & 2.9$\pm$4.2 & -4.1$\pm$2.1 & -6.9$\pm$1.3 \\ 
\end{tabular}
\caption{Indirect constraints at the one-sigma level assuming $\Lambda = 1$~TeV 
and using $\mu = m_Z$ for the renormalization scale. Fit~1 -- Fit~4 are described
in the text. \label{table-linear-indirect}}
\end{table}
The central values for $\fwww$ depend upon $m_H$ only through the SM $m_H$ 
contributions.  On the other hand, both the central values $\fw$ and $\fb$ have 
a complicated dependence on $m_H$.  Comparing Fit~2 to Fit~1, the data in 
Table~\ref{table-lep2} does not effect the magnitude of the error bars, but 
does lead to a small shift in the best-fit central values.  Proceeding to Fit~3,
the overall results for the LEP2 program are more promising.  While the error 
bars on $\fwww$ are unaffected, there is a tiny effect upon the error bars
for $\fw$, and for $\fb$ the magnitude of the error bars is, in some cases,
reduced by more than 10\%.  The improvement at a future linear collider (Fit~4) 
is significant for all three parameters.  

\begin{figure}[htb]
\begin{center}
\leavevmode\psfig{file=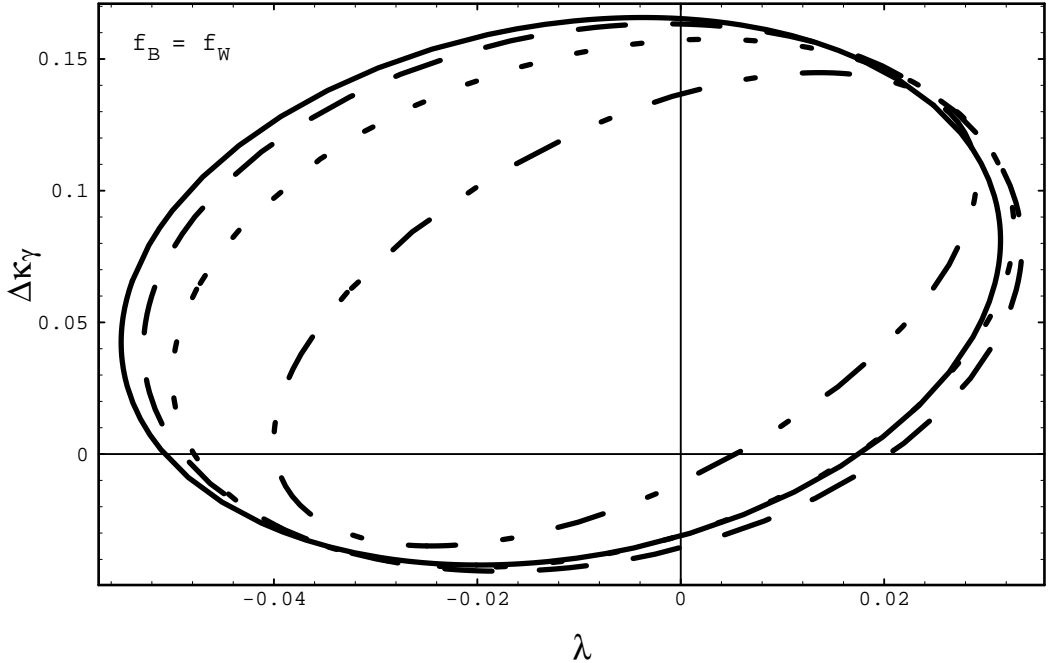,angle=0,width=14cm,silent=0}
\end{center}
\caption{Constraints at the 95\% confidence level in the $\Delta\kappa_\gamma$
{\em versus} $\lambda$ plane for $m_H = 300$~GeV subject to the constraints 
of Eqns.~{\protect\ref{hisz-relations}} with $f_B = f_W$.  The solid, dashed,
dotted and dot-dashed curves correspond to Fit~1 (low-energy and $Z$-pole data), 
Fit~2 (LEP1.5 and early LEP2 data included), Fit~3 (LEP2 included) and Fit~4
(linear collider included) respectively.  For the scale of new physics and 
the renormalization scale, $\Lambda = 1$~TeV and $\mu = m_Z$ have been employed.}
\label{fig-fw-eq-fb} 
\end{figure}

\begin{figure}[htb]
\begin{center}
\leavevmode\psfig{file=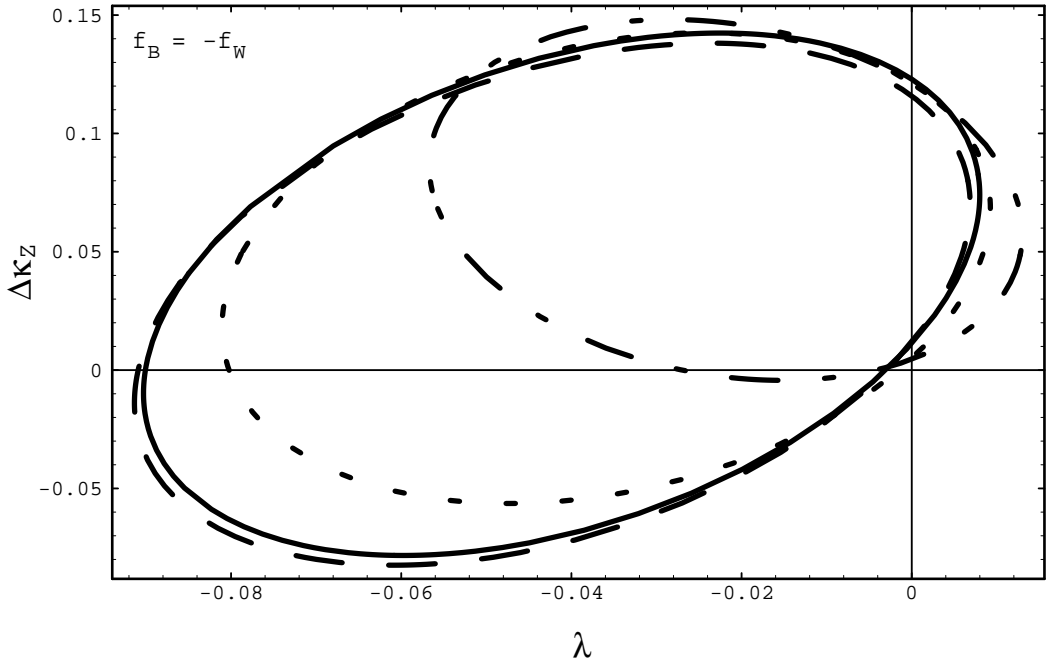,angle=0,width=14cm,silent=0}
\end{center}
\caption{Constraints at the 95\% confidence level in the $\Delta\kappa_Z$
{\em versus} $\lambda$ plane for $m_H = 300$~GeV subject to the constraints 
of Eqns.~{\protect\ref{hisz-relations}} with $f_B = -f_W$.  The solid, dashed,
dotted and dot-dashed curves correspond to Fit~1 (low-energy and $Z$-pole data), 
Fit~2 (LEP1.5 and early LEP2 data included), Fit~3 (LEP2 included) and Fit~4
(linear collider included) respectively.  For the scale of new physics and 
the renormalization scale, $\Lambda = 1$~TeV and $\mu = m_Z$ have been employed.}
\label{fig-fw-eq-minus-fb} 
\end{figure}

In Fig.~\ref{fig-fw-eq-fb},
the best fit in the 
$\Delta\kappa_\gamma$ {\em versus} $\lambda$ plane is shown for $m_H = 300$~GeV
subject to the constraints of Eqns.~\ref{hisz-relations}.  There are three 
independent variables, $\fw$, $\fb$ and $\fwww$.  A two-dimensional projection
is obtained by setting $f_B = f_W$.  In Fig.~\ref{fig-fw-eq-minus-fb}
a two dimensional projection in the $\Delta\kappa_Z$ {\em versus} $\lambda$ 
plane for $m_H = 300$~GeV has been obtained by choosing $f_B = -f_W$ (and 
therefore $\Delta\kappa_\gamma = 0$).  The bounds on $\lambda$ 
from Figure~\ref{fig-fw-eq-minus-fb} are noticeably weaker than those of 
Fig.~\ref{fig-fw-eq-fb}.  This illustrates an important aspect of these 
indirect bounds.  Without restricting the discussion to a particular model, it 
is impossible to predict relations between the various parameters, and hence the 
possibility of cancellations between them must be considered.  Furthermore, the 
central values which are obtained in the fits are sensitive to $\alpha_s$, 
the error on the hadronic contribution to the running of $\alpha_{\rm QED}$,
the top-quark mass and, as shown in Table~\ref{table-linear-indirect}, the 
unknown Higgs-boson mass.  All four fits are reflected in each of these figures.
Looking at the dashed curve we see that the LEP1.5 and available LEP2 data leads 
to only a tiny shift in the central values of the solid curves.  The dotted 
curve reflects an important reduction in the available parameter space by the 
end of LEP2, and the dot-dashed curves show serious improvement at the linear 
collider.

These indirect bounds are particularly interesting when compared to the direct 
bounds 
which may be obtained by studying $W$-boson pair production.  Such bounds,
taken from Ref.~\cite{hhis96}, are summarised in Table~\ref{table-linear-eeww}.
\begin{table}[htb]
\begin{tabular}{|c||c|c|c|}
       & $f_{WWW}$ & $f_{W}$ & $f_{B}$ \\ \hline\hline
LEP~II & 10        & 7.1     & 46      \\ \hline 
LC     & 0.23      & 0.10    & 0.25    \\
\end{tabular}
\caption{Direct one-sigma limits from $e^+e^-\rightarrow W^+W^-$ assuming 
$\Lambda = 1$TeV.  (From Ref.~{\protect\cite{hhis96}}.) In the first row are 
the constraints from LEP~II at 175GeV with ${\cal L}^{\rm int} = 
500{\rm pb}^{-1}$, and the second row contains results for a 500GeV future 
linear collider with ${\cal L}^{\rm int} = 50{\rm fb}^{-1}$.  The one-sigma 
allowed region is approximately symmetric about zero.}
\label{table-linear-eeww}
\end{table}
The numbers in this table make the same assumptions about a linear collider as 
does Fit~4.  However, for LEP2, Ref.~\cite{hhis96} considered only 
${\cal L}^{\rm int} = 
500{\rm pb}^{-1}$ at $\sqrt{s} = 175$~GeV.  If the parameters in Fit~3 more 
accurately describe the actual LEP2 program, then the numbers in 
Table~\ref{table-linear-eeww} are pessimistic.  The errors quoted are purely 
statistical, and more events should be expected with more luminosity 
plus higher energy which tends to overcome threshold suppression factors.  As
discussed above, the numbers in Table~\ref{table-linear-indirect} are somewhat 
optimistic.  By the completion of LEP2, using the numbers from the tables,
the ratios of the indirect to the 
direct bound are 1:1 for $\fwww$, 1:2 for $\fw$ and 1:5--1:21 (depending upon 
$m_H$) for $\fb$.  Taking into account the above discussion, it is safe to say 
that the direct and indirect bounds probe the same order of magnitude, and 
hence both studies should be optimized.  Proceeding to the linear collider,
the improvements in the indirect measurements are insufficient to keep pace 
with the improved direct measurements.  

Turning to the nonlinear realization of SSB, the one-sigma bounds on $\alpha_2$,
$\alpha_3$ and $\alpha_9$ are presented in Table~{table-nl-indirect}.
\begin{table}[tbh]
\begin{tabular}{|l|c|c|c|c|}
&     Fit~1        &      Fit~2       &      Fit~3       &      Fit~4      \\
\hline\hline 
$\alpha_2$ 
& 0.252$\pm$0.053  & 0.250$\pm$0.053  & 0.261$\pm$0.051  & 0.211$\pm$0.037 \\
\hline
$\alpha_3$ 
& -0.119$\pm$0.023 & -0.117$\pm$0.023 & -0.121$\pm$0.022 & -0.142$\pm$0.019 \\
\hline
$\alpha_9$ 
& -0.03$\pm$0.16   & -0.03$\pm$0.16   & -0.261$\pm$0.092 & -0.242$\pm$0.040 \\
\end{tabular}
\caption{Indirect constraints at the one-sigma level assuming $\Lambda = 1$~TeV 
and using $\mu = m_Z$ for the renormalization scale. Fit~1 -- Fit~4 are described
in the text. \label{table-nl-indirect}}
\end{table}
The inclusion of LEP1.5 and early LEP2 data (Fit~2) has a small effect on the 
best-fit cental values, but there is no change in the error bars.  The complete 
LEP2 program makes a tiny reduction in the error bars for $\alpha_2$ and 
$\alpha_3$, but the error on $\alpha_9$ is reduced by 40\%.  At the linear 
collider significant improvements are achieved for $\alpha_2$ and $\alpha_9$, 
and the error on $\alpha_3$ is also reduced.  As in the discussion following 
Table~\ref{table-linear-indirect}, these bounds are weakened by correlations 
among the various parameters and by uncertainties in SM parameters.

The corresponding direct constraints from Ref.~\cite{hhis96} are presented in 
Table~\ref{table-nl-eeww}.
\begin{table}[htb]
\begin{tabular}{|c||c|c|c|}
       & $\alpha_2$ &  $\alpha_3$ &  $\alpha_9$ \\ \hline\hline
LEP~II & 0.34       & 0.053       & 0.10        \\ \hline 
LC     & 0.0018     & 0.00072     & 0.00078 
\\ 
\end{tabular}
\caption{Direct one-sigma limits from $e^+e^-\rightarrow W^+W^-$ assuming
$\Lambda = 1$~TeV.  (From Ref.~{\protect\cite{hhis96}}.) In the first row are 
the constraints from LEP~II at 175GeV with ${\cal L}^{\rm int} = 
500{\rm pb}^{-1}$, and the second row contains results for a 500GeV future 
linear collider with ${\cal L}^{\rm int} = 50{\rm fb}^{-1}$.  The one-sigma 
region is approximately symmetric about zero.}
\label{table-nl-eeww}
\end{table}
Again, the assumptions in obtaining these table entries are pessimistic 
compared to the numbers used in Fit~3.
The post-LEP2 ratios of indirect to direct bounds are 1:7 for $\alpha_2$, 
1:2.5 for $\alpha_3$ and 1:1 for $\alpha_9$.  Again, the direct and indirect 
bounds are of the same order at LEP2.  At the linear collider the direct 
constraints are more than an order of magnitude better than the indirect. 

\begin{figure}[htb]
\begin{center}
\leavevmode\psfig{file=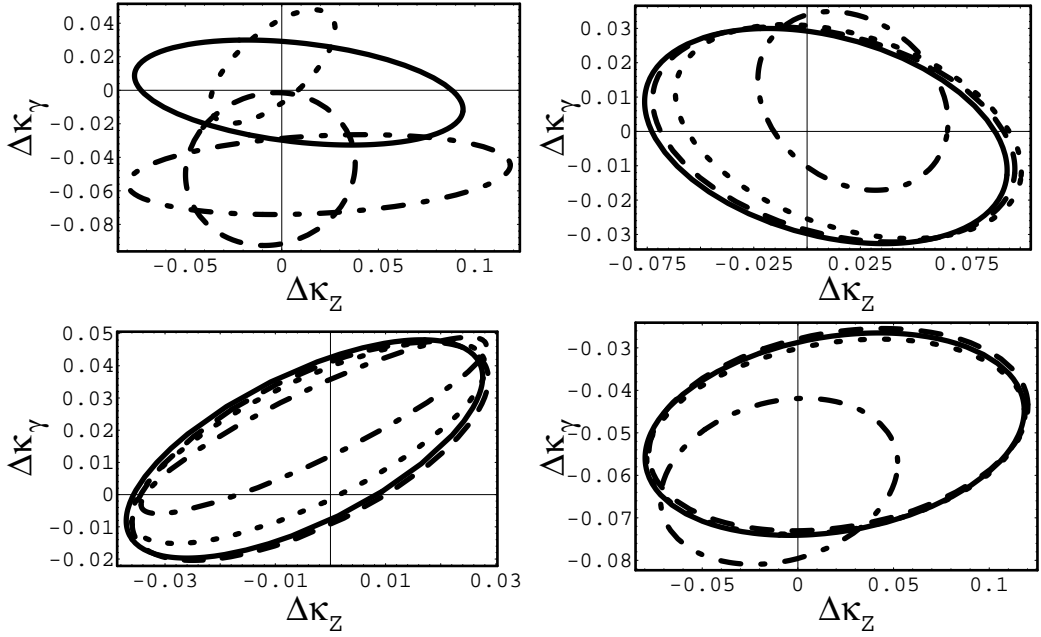,angle=0,width=14cm,silent=0}
\end{center}
\caption{Constraints at the 95\% confidence level in the 
$\Delta\kappa_\gamma$--$\Delta\kappa_Z$ plane.  Fig.~(a)
compares the linear realization of SSB (with $\fwww$ = 0) for $m_H = 100$~GeV
(solid curve), $m_H = 300$~GeV (dashed curve) and $m_H = 700$~GeV (dotted 
curve) with the nonlinear realization of SSB with $\alpha_9 = 0$ (dot-dashed 
curve).  Then, Fig.~(b) and (c) specialize to $m_H = 100$~GeV and 
$m_H = 700$~GeV, respectively, and Fig.~(d) concerns the nonlinear realization 
of SSB.  In Fig.~(b)-(d), the solid, dashed, dotted and dot-dashed curves 
correspond to Fit~1, Fit~2,Fit~3 and Fit~4 respectively.  $\Lambda = 1$~TeV and 
$\mu = m_Z$ have been assumed.}
\label{fig-dkg-vs-dkz} 
\end{figure}

Finally, Fig.~\ref{fig-dkg-vs-dkz} 
is a projection in the 
$\Delta\kappa_\gamma$--$\Delta\kappa_Z$ plane.  Fig.~\ref{fig-dkg-vs-dkz}(a)
compares the linear realization of SSB (with $\fwww$ = 0) for $m_H = 100$~GeV
(solid curve), $m_H = 300$~GeV (dashed curve) and $m_H = 700$~GeV (dotted 
curve) with the nonlinear realization of SSB with $\alpha_9 = 0$ (dot-dashed 
curve).  Then, Fig.~\ref{fig-dkg-vs-dkz}(b) and (c) specialize to $m_H = 100$~GeV
and $m_H = 700$~GeV, respectively, and Fig.~\ref{fig-dkg-vs-dkz}(d) concerns 
the nonlinear realization of SSB.  In Fig.~\ref{fig-dkg-vs-dkz}(b)-(d), the 
solid, dashed, dotted and dot-dashed curves correspond to Fit~1, Fit~2, Fit~3 
and Fit~4 respectively.  All fits are at the 95\% confidence level.

Examining Fig.~\ref{fig-dkg-vs-dkz}(a), the contours for the linear realization 
of SSB are highly dependent upon $m_H$, and, while all are consistent with SM,
the intermediate mass of $m_H = 300$~GeV disfavors the point 
$(\Delta\kappa_\gamma,\Delta\kappa_Z) = (0,0)$.  The allowed parameter space 
clearly decreases with increasing $m_H$.  On the other hand, the contour 
for the nonlinear realization is clearly inconsistent with the SM.  
Proceeding to Fig.~\ref{fig-dkg-vs-dkz}(a) and (b), modest reductions in 
the allowed regions may be achieved at LEP2, while the improvements at the 
linear collider are much more significant.  For the nonlinear realization of 
SSB, Fig.~\ref{fig-dkg-vs-dkz}(d) shows minimal gains at LEP2, but the 
improvement at the linear collider is dramatic.


\section{Conclusions}

Nonstandard $WW\gamma$ and $WWZ$ vertices may be probed directly \via\/
their tree-level contributions to processes such as $e^+e^-\rightarrow W^+W^-$
or indirectly \via\/ their loop-level contributions to electroweak observables.
The direct constraints on these couplings will improve significantly as data 
is accumulated at the ongoing LEP2 experiments.  The resulting direct constraints
and the current indirect constraints, allowing for theoretical uncertainties 
in the latter, probe new physics at roughly the same level.  Therefore it is 
desirable to also improve the indirect constraints as much as possible.  The 
LEP1.5 and LEP2 data which is presently available leads to only a very tiny 
improvement, but it is anticipated that significant gains will have been 
accomplished before the end of the LEP2 experiments.

Truly significant improvements in the loop-level constraints are expected 
from the inclusion of data collected at a future linear collider operating 
at $\sqrt{s} = 500$~GeV.  However, the improvements in the direct constraints 
through the study of $W$-boson pair production at the same facility will be 
much more impressive, and it is very that the better measurements 
will be obtained from these direct studies.


\section*{Acknowledgements}
Discussions with Kaoru Hagiwara are gratefully acknowledged.  The author is 
is thankful to Seiji Matsumoto, Sally Dawson and Sher Alam for prior 
collaborations on related topics.  This work was 
supported in part by the National Science Foundation (NSF) through grant no.
INT9600243, and in part by the Japan Society for the Promotion of Science
(JSPS).  


\end{document}